\documentclass{ptephy}

\usepackage{bm,amssymb}

\graphicspath{{./Figures/}}

\begin{document}
\newcommand{\diff}{\mathrm{d}}
\newcommand{\p}{\partial}
\newcommand{\e}{\varepsilon}
\newcommand{\Diff}{{\mathcal{D}}}
\newcommand{\V}{{\mathcal{V}}}
\newcommand{\up}{\uparrow}
\newcommand{\down}{\downarrow}
\newcommand{\be}{\begin{equation}}      
\newcommand{\ee}{\end{equation}}      
\newcommand{\bea}{\begin{eqnarray}}      
\newcommand{\eea}{\end{eqnarray}}    

\title{Fermionic Functional Renormalization Group Approach to Bose-Einstein Condensation of Dimers}

\author{\name{\fname{Yuya} \surname{Tanizaki}}{1}}

\address{\affil{1}{Department of Physics, The University of Tokyo, Tokyo 113-0033, Japan}
\affil{1}{Theoretical Research Division, Nishina Center, RIKEN, Wako 351-0198, Japan}
\email{yuya.tanizaki@riken.jp}}

\begin{abstract}%
Fermionic functional renormalization group (f-FRG) is applied to describe Bose-Einstein condensation (BEC) of dimers for a two-component fermionic system with attractive contact interaction. 
In order to describe the system of dimers without introducing auxiliary bosonic fields (bosonization), we propose a new exact evolution-equation of the effective action in f-FRG with an infrared regulator for the fermion vertex. Then we analyze its basic properties in details. 
We show explicitly that the critical temperature of the free Bose gas is obtained naturally by this method without bosonization. Methods to make systematic improvement from the deep BEC limit are briefly discussed. 
\end{abstract}

\subjectindex{A63, B32, B36}

\maketitle
\section{Introduction}\label{sec:intro}
Recent developments in the field of ultracold atomic gases continue to stimulate various areas of physics. Since the Feshbach resonance technique enables to tune the inter-atomic interactions with a high controllability, one can experimentally simulate quantum systems in a very idealistic situation. 
Especially, it is important to point out that ultracold atomic gases provide a realization of the Bardeen-Cooper-Schrieffer (BCS) to Bose-Einstein condensation (BEC) crossover for a two-component fermionic system with attractive interaction \cite{greiner2003emergence,PhysRevLett.91.250401,PhysRevLett.92.040403,PhysRevLett.92.120403}, as theoretically predicted from the mean-field type calculations \cite{PhysRev.186.456,leggett1980diatomic,nozieres1985bose}. 

The BCS-BEC crossover is a crossover phenomenon from the BCS superfluid for weakly-coupled Fermi liquid to the BEC for tightly-bound composite bosons as the attraction between fermions becomes stronger.  
In order to reveal thermodynamic properties of this crossover phenomenon, various kinds of non-perturbative analysis of quantum field theory have been developed, including Monte-Carlo simulations  \cite{PhysRevLett.96.160402,PhysRevLett.101.090402,PhysRevLett.103.210403,PhysRevA.82.053621,PhysRevB.76.165116}, the $\e$-expansion \cite{PhysRevA.75.063618}, the functional renormalization group approach with auxiliary bosonic field \cite{birse2005pairing,PhysRevB.80.104514,PhysRevB.78.174528,ANDP:ANDP201010458,PhysRevA.81.063619}, and the self-consistent $t$-matrix method \cite{haussmann1993crossover,PhysRevB.49.12975,PhysRevA.75.023610,arXiv:1109.2307}. 

Functional renormalization group (FRG) \cite{wetterich1993exact,Morris1,ellwanger1994flow} is a non-perturbative tool in quantum field theory. By introducing appropriate bosonic fields, FRG provides systematic studies of the second-order phase transition. 
Since both the large-size Cooper pairs in the BCS limit and the tightly bound dimers in the BEC limit in the BCS-BEC crossover behave as effective bosonic fields, FRG with auxiliary field method has been widely applied to describe the BCS-BEC crossover. 
However, the auxiliary field method requires some knowledge of the ground state property of the system, so that the choice of the auxiliary fields is not  known \textit{a priori}.

The purpose of this paper is to develop fermionic FRG (f-FRG) to study the BCS-BEC crossover: An advantage of f-FRG is to provide us with systematic and unbiased analysis of interacting fermionic systems \cite{Salmhofer:2001tr,gersch2008superconductivity,RevModPhys.84.299}. In our previous paper \cite{Tanizaki:2013doa,Tanizaki:2013fba}, the BCS theory and its Gorkov and Melik-Barkhudarov correction are reproduced within f-FRG. Furthermore, the effect of the self-energy correction to the f-FRG flow equation is discussed in details in the BCS side. 
In this paper, we will consider the opposite side, i.e. the BEC side of the BCS-BEC crossover, within the f-FRG formalism. This is a non-trivial problem, since low-energy excitations are one-particle excitations of composite bosons, while fermionic excitations are highly suppressed due to the binding energy. 
Therefore, naive f-FRG formalism without auxiliary bosons is not suitable to describe the BEC limit, because it only suppresses fermionic excitations by introducing an IR regulator in the fermion propagator. 

In this paper, we propose a new evolution equation of f-FRG in the BEC regime by introducing an IR regulator in the four-fermion vertex (vertex IR regulator). 
The structure of the new flow equation for a non-relativistic fermionic system is studied in details. In the limit where the dimer density is low enough, the coupled flow equations for the four-point vertex and the fermion self-energy are solved, and the critical temperature of the free Bose gas is reproduced without bosonization. 
Methods to go beyond the free gas limit in a systematic way are also discussed.  

The organization of this paper is  as follows. 
In Sec.\ref{sec:mfa}, we will find an appropriate starting point of f-FRG describing the BEC of composite bosons. After defining a fermionic field theory for the two-component fermionic system with attractive contact interaction, its equivalent bosonic description is reviewed to find out low-energy degrees of freedom in the BEC limit.  Introducing an IR regulator in a bosonic propagator in the BEC limit, we derive a vertex IR regulator for the corresponding fermionic field theory. 
In Sec.\ref{sec:ffrg}, we first develop a FRG formalism with the vertex IR regulator and derive its flow equation for the self-energy and the four-point vertex function in Sec.\ref{subsec:frg_quartic}. Since we are interested in the low-density system, we solve the flow equation for the four-point vertex function in the vacuum in Sec.\ref{subsec:flow_vac}. The self-energy correction is very important to take into account the existence of composite bosons in evaluating the particle number density. Therefore, in Sec.\ref{subsec:flow_se} we will derive an approximate solution for the self-energy flow equation of our f-FRG formalism in order to derive the BEC critical temperature $T_c$ of free Bose gas. In Sec.\ref{subsec:improv}, methods to make systematic improvement from the deep BEC limit are briefly discussed. 
In Sec.\ref{sec:summary}, we summarize our result and also give concluding remarks.

\section{Derivation of the vertex IR regulator}\label{sec:mfa}

Let us consider the non-relativistic two-component fermions $\psi=\left(\begin{array}{c} \psi_{\up}\\\psi_{\down}\end{array}\right)$ with a contact interaction: 
\be
S_f[\overline{\psi},\psi]=\int_0^{\beta}\diff \tau\int \diff^3\bm{x}
\left[\overline{\psi}\left(\p_{\tau}-{\nabla^2\over 2m}-\mu\right)\psi(x)
+g\overline{\psi}_{\up}\overline{\psi}_{\down}\psi_{\down}\psi_{\up}(x)\right],
\label{mfa01}
\ee
with $\beta(=1/T)$ the inverse temperature, $\mu$ the chemical potential, $m$ the fermion mass, and $g$ the bare coupling constant. The ultraviolet (UV) regularized form of (\ref{mfa01}) in the momentum space is given by 
\be
S_f[\overline{\psi},\psi]=\int_p^{(T)} \overline{\psi}_{p} G^{-1}(p) \psi_{p}
+g\int_p^{(T)} e^{-ip^0 0^+}\int_{q,q'\le \Lambda}^{(T)}
\overline{\psi}_{\up,p/2+q}\overline{\psi}_{\down,{p}/2-q}
\psi_{\down,{p}/2-q'}\psi_{\up,p/2+q'},
\label{mfa02}
\ee
where $G^{-1}(p)=i p^0+\bm{p}^2/2m-\mu$ with $p=(p^0,\bm{p})$, $\psi_{\sigma,p}$ denotes Fourier coefficient of $\psi_{\sigma}(x)$, $\int_p^{(T)}=\int{\diff^3\bm{p}\over (2\pi)^3}{1\over \beta}\sum_{p^0}$, and ``$\le \Lambda$'' in the relative momentum integration denotes the spatial momentum UV cutoff for $\bm{q}$ and $\bm{q}'$. 
Renormalization condition can be denoted for the bare coupling $g$ using the scattering length $a_s$ as 
$g^{-1}={m\over 4\pi a_s}-{m \Lambda\over 2\pi^2}$. 
Since our interest is in the BEC limit, we will concentrate on analyzing the physics when $a_s$ is positive: $a_s>0$. 

In order to set up a new f-FRG formalism applicable to the BEC side of the BEC-BCS crossover, we need to find an appropriate infrared (IR) regulator to control the tightly bound fermion pairs (dimers).  For the purpose to find such an IR regulator, we will first review, in Sec.\ref{subsec:mfa},  the derivation of the critical temperature $T_c$ for the free Bose gas starting from the bosonized version of (\ref{mfa01}).  This enables us to identify the correct low-energy degrees of freedom to be regulated in the bosonized theory, and hence in the original fermionic theory. With such knowledge,  we introduce, in Sec.\ref{subsec:IRreg}, a vertex IR regulator which plays a key role to control the dimer excitations of the original fermionic theory.        

\subsection{Brief review on $T_c$ in the deep BEC limit}\label{subsec:mfa}
In order to find out low-energy degrees of freedom in the BEC side, we give a brief review of the calculation of $T_c$ within the low-density approximation of dimers. For this purpose, we once introduce an equivalent bosonic description of the fermionic field theory defined in (\ref{mfa02}). 

Using the Hubbard-Stratonovich transformation \cite{Stratonovich1958,PhysRevLett.3.77}, we can introduce the bosonic auxiliary field $\Delta(x)$ to rewrite (\ref{mfa02}) as 
\bea
S_{b+f}[\overline{\psi},\psi,\Delta^*,\Delta]&=&\int_p^{(T)} \overline{\psi}_p G^{-1}(p)\psi_p-{1\over g}\int_p^{(T)} e^{-i p^0 0^+}\Delta^*_p\Delta_p \nonumber\\
&+&\int_p^{(T)}e^{- ip^0 0^+}\left(\Delta^*_p\int_{q\le \Lambda}^{(T)}\psi_{\down,p/2-q}\psi_{\up,p/2+q}+\mathrm{h.c.}\right), 
\label{mfa04}
\eea
where ``$\mathrm{h.c.}$'' denotes the Hermitian conjugate of the first term in the large parenthesis. 
By integrating out fermions, we can find the bosonic action $S_b[\Delta^*,\Delta]$:
\be
S_b[\Delta^*,\Delta]=\int_p^{(T)}\Gamma_b^{(2)}(p)\Delta^*_p\Delta_p+{1\over 2^2}\int_{p,q,q'}^{(T)}\Gamma_b^{(4)}(p;q,q')\Delta^*_{p/2+q}\Delta^*_{p/2-q}\Delta_{p/2-q'}\Delta_{p/2+q'}+\cdots. 
\label{mfa05}
\ee
The explicit formula for $\Gamma_b^{(2)}$ is given as 
\be
\Gamma_b^{(2)}(p)=-\left\{{m\over 4\pi a_s}+\int{\diff^3\bm{q}\over (2\pi)^3}\left[{1-n_F({(\bm{p}/2-\bm{q})^2\over 2m}-\mu)-n_F({(\bm{p}/2+\bm{q})^2\over 2m}-\mu)\over \bm{q}^2/m+(i p^0+\bm{p}^2/4m-2\mu)}-{1\over \bm{q}^2/m}\right]\right\},  
\label{mfa06}
\ee
with $n_F$ the Fermi-Dirac distribution function. 
Since the self-energy correction to the bosonic classical propagator (\ref{mfa06}) becomes smaller in the low-density limit, the critical point of the superfluid phase transition is given by $\Gamma_b^{(2)}(p=0)=0$ when the density $n$ is small enough: $(n^{1/3}a_s)^{-1}\to +\infty$. Since $a_s>0$, the fermionic chemical potential $\mu$ must be negative for this condition. Up to an exponentially small correction in terms of $\beta/(m a_s^2)$, this requires that $\mu=-{1/( 2m a_s^2)}$, 
which is exactly half of the binding energy of the composite boson in the vacuum. Within the same order approximation, $\Gamma_b^{(2)}(p)$ can be replaced by the bosonic propagator in the vacuum: 
\be
\Gamma_b^{(2)}(p)\simeq \left.\Gamma_b^{(2)}(p)\right|_{T=0,\mu=-1/2ma_s^2-0^+}={m\over 4\pi a_s}\left(\sqrt{1+ma_s^2\left(i p^0+{\bm{p^2}\over 4m}\right)}-1\right). 
\label{mfa08}
\ee
With this approximation, we can indeed obtain the BEC transition temperature of free Bose gas: $T_c/\e_F=\left({2/( 9\pi \zeta(3/2)^2)}\right)^{1/3}=0.218\ldots$ with $\e_F=(3\pi^2 n)^{2/3}/(2m)$ and $n$ the particle number density. 

\subsection{Vertex IR regulator in f-FRG}\label{subsec:IRreg}
We can now give a definition of a vertex IR regulator for the fermionic field theory to study the BEC side of the BCS-BEC crossover. 
In order to find an appropriate starting point of our new formalism of f-FRG which can describe the deep BEC limit, we need to regulate low-energy one-particle excitations of dimers. Obeying the general strategy of FRG, we add the IR regulator $\delta S_{b,k}[\Delta^*,\Delta]$ to $S_{b+f}$ in (\ref{mfa04}), which suppresses low-energy dimer excitations with excitation energy $\lesssim k^2/4m$: 
\be
\delta S_{b,k}[\Delta^*,\Delta]=\int_p^{(T)} R_k(\bm{p})\Delta^*_p\Delta_p. 
\label{mfa11}
\ee
A typical example of the IR regulating function $R_k$ is Litim's optimized regulator \cite{litim2000optimisation}: 
\be
R_k(\bm{p})={m^2 a_s\over 8\pi}\left({k^2\over 4m}-{\bm{p}^2\over 4m}\right)\theta(k^2-\bm{p}^2). 
\label{mfa12}
\ee
By integrating out the bosonic degrees of freedom from $S_{b+f}[\overline{\psi},\psi,\Delta^*,\Delta]+\delta S_{b,k}[\Delta^*,\Delta]$, we can find the IR regulating term $\delta S_{f,k}[\overline{\psi},\psi]$ in the original fermionic field theory: $\exp\left(-[S_{f}+\delta S_{f,k}]\right)=\int\Diff\Delta^*\Diff\Delta\exp\left(-[S_{b+f}+\delta S_{b,k}]\right)$. 
This functional integration can be done exactly, and $\delta S_{f,k}$ is obtained as a quartic term instead of the frequently used two-point IR regulating function: 
\be
\delta S_{f,k}[\overline{\psi},\psi]=\int_p^{(T)} {g^2 R_k(\bm{p})\over 1-g R_k(\bm{p})}e^{-ip^0 0^+}
\int_{q,q'\le \Lambda}^{(T)}\overline{\psi}_{\up,p/2+q}\overline{\psi}_{\down,{p}/2-q}\psi_{\down,{p}/2-q'}\psi_{\up,p/2+q'}. 
\label{mfa13}
\ee
This is an explicit definition for our vertex IR regulator of f-FRG. 
We will denote the coefficient in (\ref{mfa13}) as $g_k(p)$: $g_k(p)=g^2R_k(\bm{p})/(1-g R_k(\bm{p}))$.  
In the limit $k\to \infty$, the IR regulator $\delta S_{f,k}$ is the negative of the interaction part of the classical action $S_f$ since $g_{k=\infty}(p)=-g$, and thus the system becomes free theory when $k=\infty$. 
This argument suggests us to use a vertex IR regulator in order to suppress low-energy excitations in the BEC limit within f-FRG formalism, and the flow starts from the free fermionic theory and converges to the weakly interacting bosonic theory.

\section{Fermionic FRG formalism for bosonic fluctuations}\label{sec:ffrg}
Motivated by the analysis in Sec.\ref{subsec:IRreg}, we first derive a new exact evolution equation of f-FRG with a vertex IR regulator in this section. Structure of the flow equation is studied in a general way, and the new formalism of f-FRG is applied to describe the BEC of dimers without bosonization. Since we are interested in a low-density system, the flow of the four-point vertex is calculated in the vacuum limit, and the self-energy flow is solved within a first order approximation in terms of the density. 

\subsection{FRG with a vertex IR regulator}\label{subsec:frg_quartic}
In this subsection, we describe the general formalism of FRG with the vertex IR regulator, not specific to a fermionic field theory. 
We consider a field theory with a classical action $S[\phi]$ of a field $\phi$, which contains a bare propagator and a four-point vertex: 
\be
S[\phi]={1\over 2}\phi_{\alpha_1}G^{-1,\alpha_1\alpha_2}\phi_{\alpha_2}+{1\over 4!}g^{\alpha_1\alpha_2\alpha_3\alpha_4}\phi_{\alpha_1}\phi_{\alpha_2}\phi_{\alpha_3}\phi_{\alpha_4}, 
\ee
where $\alpha_i$ denotes the label to be summed up, such as space-time coordinates, particle species, and internal degrees of freedom. We add a vertex IR regulating term $\delta S_k$, which depends on a parameter $k$ smoothly, to this action: 
\be
\delta S_k[\phi]={1\over 4!}g_k^{\alpha_1\alpha_2\alpha_3\alpha_4}\phi_{\alpha_1}\phi_{\alpha_2}\phi_{\alpha_3}\phi_{\alpha_4}, 
\label{frg01}
\ee
with the boundary conditions $g_{k=\infty}=-g$ and $g_{k=0}=0$. 
The $k$-dependent Schwinger functional $W_k[J]$ is defined as 
\be
\exp(W_k[J])=\int\Diff\phi \exp\left(-\left(S[\phi]+\delta S_k[\phi]\right)+J^{\alpha}\phi_{\alpha}\right)
\label{frg02}
\ee
with $J$ the external source. The flow of $W_k[J]$ is determined by 
\be
\p_k W_k[J]=-\exp\left({-W_k[J]}\right)(\p_k \delta S_k)\left[{\delta_L\over \delta J}\right]\exp{W_k[J]}, 
\label{frg03}
\ee
where $\delta_L/\delta J$ denotes the left-derivative in terms of $J$. Substitution of (\ref{frg01}) into (\ref{frg03}) gives
\bea
\p_k W_k[J]&=&-{1\over 4!}\p_k g_k^{\alpha_1\alpha_2\alpha_3\alpha_4}\left({\delta_L W_k\over \delta J^{\alpha_1}}{\delta_L W_k\over \delta J^{\alpha_2}}{\delta_L W_k\over \delta J^{\alpha_3}}{\delta_L W_k\over \delta J^{\alpha_4}}\right.\nonumber\\
&+&6{\delta_L^2W_k\over \delta J^{\alpha_1}\delta J^{\alpha_2}}{\delta_L W_k\over \delta J^{\alpha_3}}{\delta_L W_k\over \delta J^{\alpha_4}}+3{\delta_L^2 W_k\over \delta J^{\alpha_1}\delta J^{\alpha_2}}{\delta_L^2 W_k\over \delta J^{\alpha_3}\delta J^{\alpha_4}}\nonumber\\
&+&\left.4{\delta_L^3 W_k\over \delta J^{\alpha_1}\delta J^{\alpha_2}\delta J^{\alpha_3}}{\delta_L W_k\over \delta J^{\alpha_4}}+{\delta_L^4 W_k\over \delta J^{\alpha_1}\delta J^{\alpha_2}\delta J^{\alpha_3}\delta J^{\alpha_4}}\right). 
\label{frg04}
\eea
We define  the $k$-dependent one-particle-irreducible (1PI) effective action $\Gamma_k[\varphi]$ as the Legendre transformation 
\be
\Gamma_k[\varphi]=J_k^{\alpha}[\varphi]\varphi_{\alpha}-W_k[J_k[\varphi]], 
\label{frg05}
\ee
where $J_k[\varphi]$ is defined as the inverse function of $\delta_L W_k[J]/\delta J=\varphi$. This is the generating functional of 1PI vertex functions at scale $k$. 
Since $\delta S_{k=0}[\phi]=0$, $\Gamma_k[\varphi]$ converges to the 1PI effective action of the original theory at $k=0$. On the other hand, since $\delta S_{k=\infty}[\phi]$ is the negative of the interaction term in $S[\phi]$, the flow of FRG with the vertex IR regulator starts from the free theory: 
\be
\Gamma_{k=\infty}[\varphi]={1\over 2}\varphi_{\alpha_1}G^{-1,\alpha_1\alpha_2}\varphi_{\alpha_2}. 
\ee

\begin{figure}[t]
\bea
\p_k\parbox{4em}{\includegraphics[width=4em]{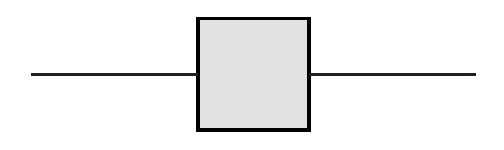}}&=&
\parbox{4em}{\includegraphics[width=4em]{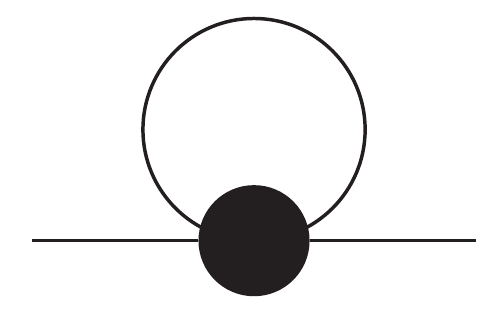}}+\parbox{6em}{\includegraphics[width=6em]{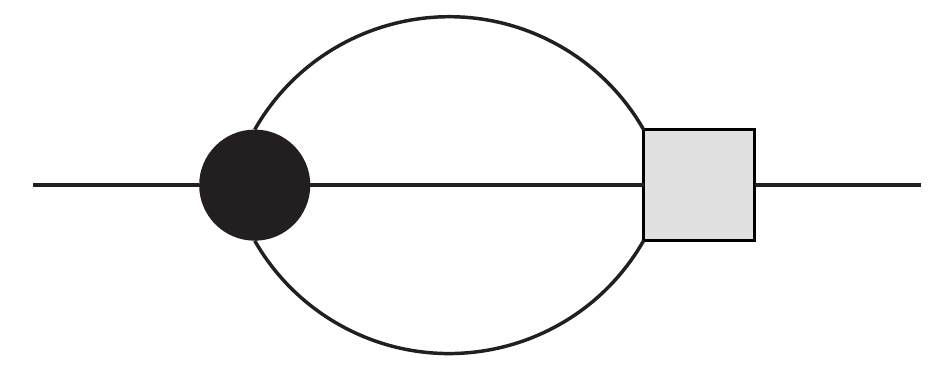}}+\parbox{7em}{\includegraphics[width=7em]{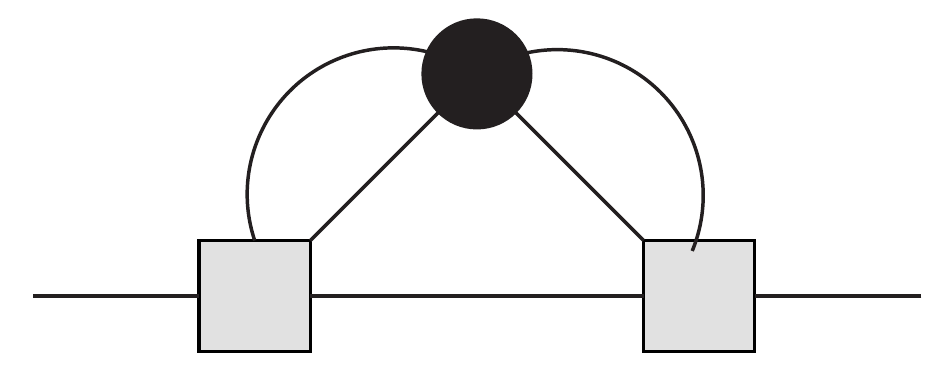}}+\parbox{3em}{\includegraphics[width=3em]{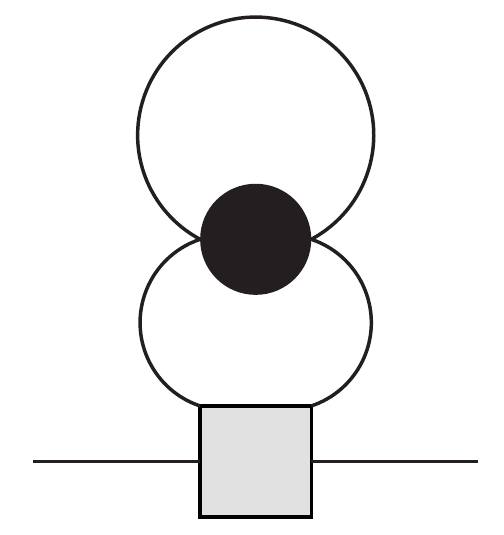}}+\parbox{4em}{\includegraphics[width=4em]{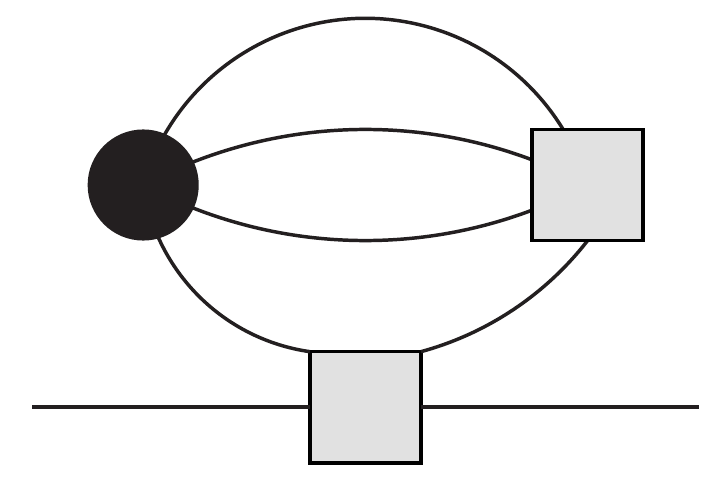}}
\nonumber
\eea
\caption{Flow equation of the self-energy $\Sigma_k$ within fourth-order vertex expansion. The square boxes represent 1PI vertex functions $\Gamma_k^{(4)}$, and the blobs denote $\p_k g_k$. Each line represents the dressed propagator $(G^{-1}-\Sigma_k)^{-1}$ at the scale $k$. }
\label{fig:flow_se}
\end{figure}

\begin{figure}[t]
\bea
\p_k\parbox{2.5em}{\includegraphics[width=2.5em]{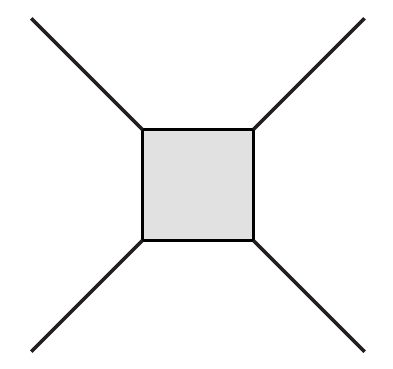}}&=&
\parbox{3em}{\includegraphics[width=3em]{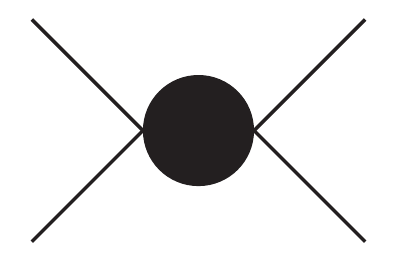}}+\parbox{5.5em}{\includegraphics[width=5.5em]{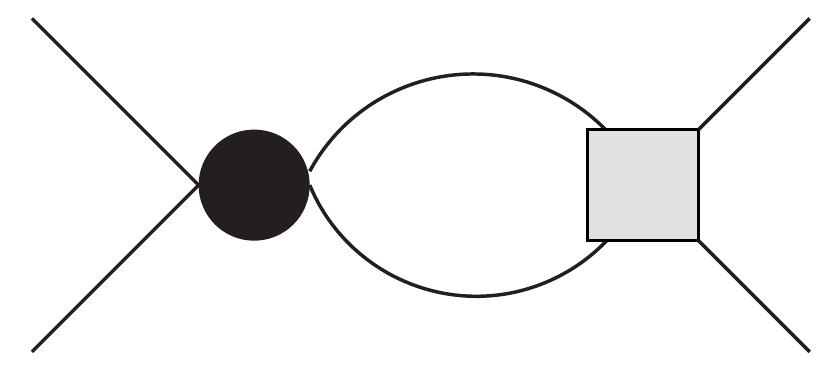}}+\parbox{8em}{\includegraphics[width=8em]{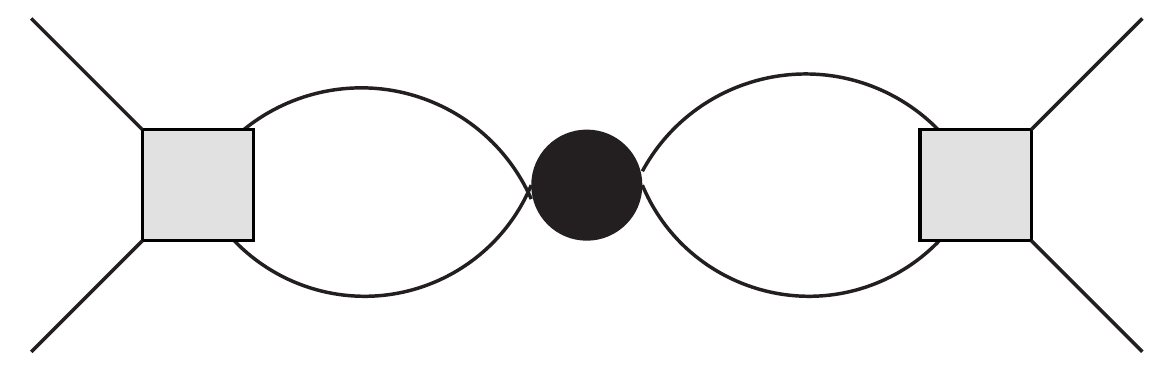}}+\parbox{5em}{\includegraphics[width=5em]{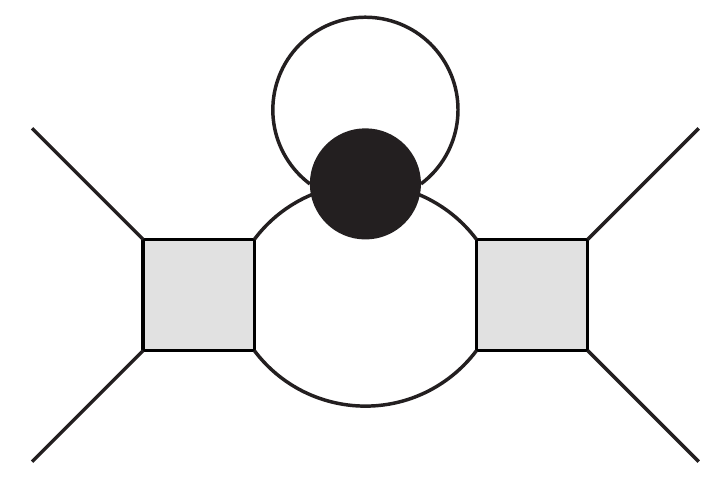}}+\parbox{5em}{\includegraphics[width=5em]{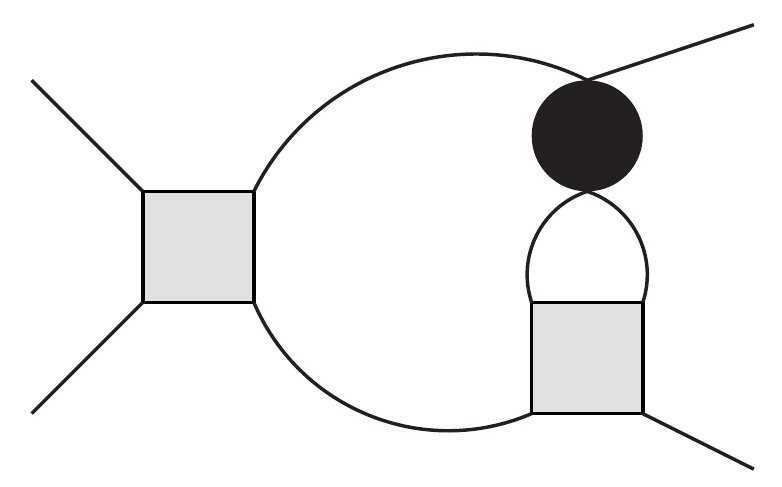}}\nonumber\\
&+&\parbox{4.5em}{\includegraphics[width=4.5em]{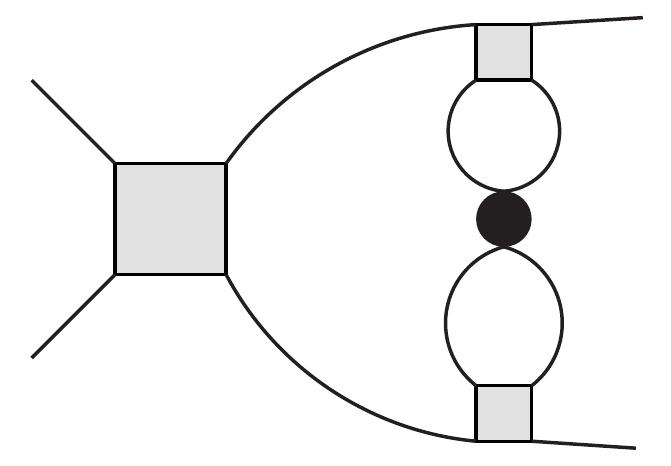}}+\parbox{6em}{\includegraphics[width=6em]{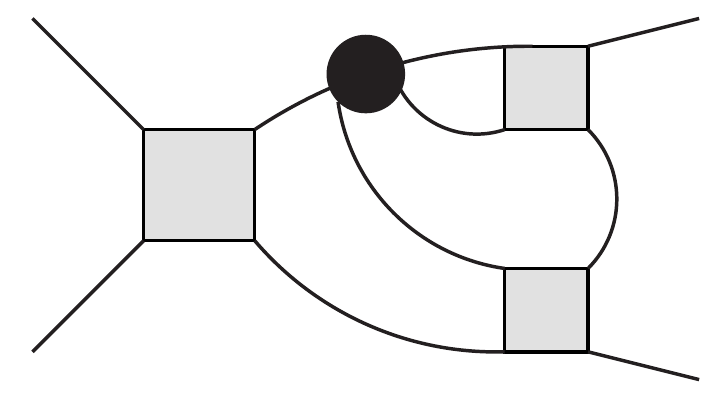}}+\parbox{7em}{\includegraphics[width=7em]{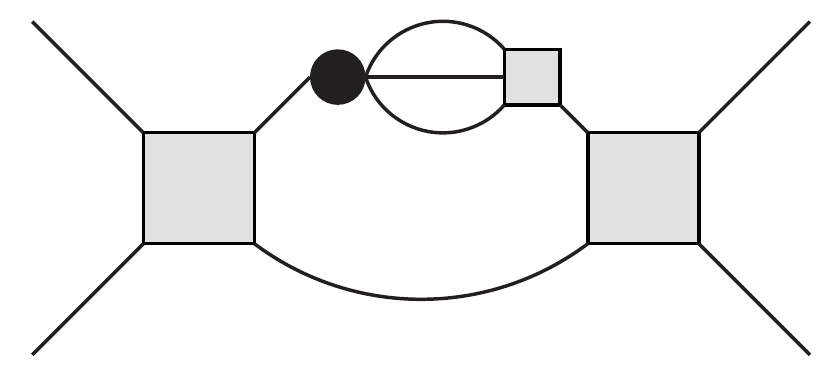}}+\parbox{7em}{\includegraphics[width=7em]{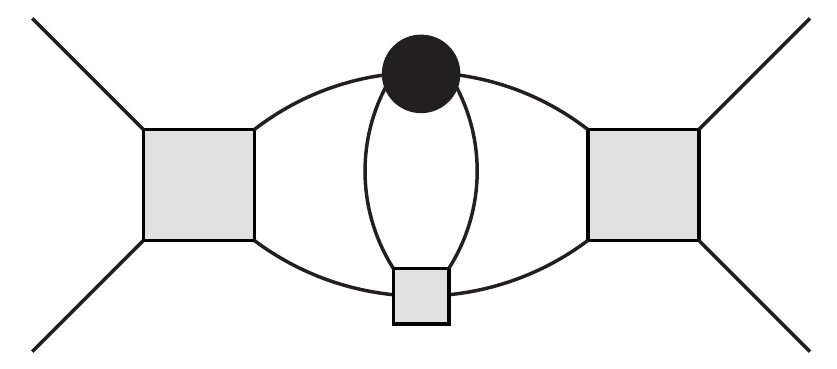}}
\nonumber
\eea
\caption{Flow equation of the four-point 1PI vertex function $\Gamma_k^{(4)}$ within fourth-order vertex expansion. The symbols are the same in Fig.\ref{fig:flow_se}. }
\label{fig:flow_4pt}
\end{figure}

Since the general property of the Legendre transformation (\ref{frg05}) indicates that $\p_k \Gamma_k[\varphi]=-(\p_k W_k)[J_k[\varphi]]$, the flow equation of $\Gamma_k[\varphi]$ is determined as 
\bea
\p_k \Gamma_k[\varphi]&=&{1\over 4!}\p_k g_k^{\alpha_1\alpha_2\alpha_3\alpha_4}
\Big(\varphi_{\alpha_1}\varphi_{\alpha_2}\varphi_{\alpha_3}\varphi_{\alpha_4}
+6\varphi_{\alpha_1}\varphi_{\alpha_2}G_{k,\alpha_3\alpha_4}
+3G_{k,\alpha_1\alpha_2}G_{k,\alpha_3\alpha_4}\nonumber\\
&+&4\varphi_{\alpha_1}G_{k,\alpha_2\beta_2}G_{k,\alpha_3\beta_3}G_{k,\alpha_4\beta_4}\Gamma_k^{(3),\beta_2\beta_3\beta_4}
+G_{k,\alpha_1\beta_1}G_{k,\alpha_2\beta_2}G_{k,\alpha_3\beta_3}G_{k,\alpha_4\beta_4}\Gamma_k^{(4),\beta_1\beta_2\beta_3\beta_4}
\nonumber\\
&+&3G_{k,\alpha_1\beta_1}G_{k,\alpha_2\beta_2}G_{k,\alpha_3\beta_3}G_{k,\alpha_4\beta_4}G_{k,\gamma_1\gamma_2}\Gamma_k^{(3),\beta_1\beta_2\gamma_1}\Gamma_k^{(3),\gamma_2\beta_3\beta_4}\Big),
\label{frg06}
\eea
with $G_k[\varphi]=\left({\delta_L\over \delta\varphi}{\delta_R\over \delta \varphi}\Gamma_k[\varphi]\right)^{-1}$ the field dependent propagator, and $\Gamma_k^{(n)}[\varphi]$ the $n$-th functional derivative of $\Gamma_k[\varphi]$. 
By taking the vertex expansion of the 1PI effective action $\Gamma_k[\varphi]$ in terms of fields $\varphi$ in (\ref{frg06}), we can obtain the flow equation for each 1PI vertex function. Expanding $\Gamma_k[\varphi]$ up to fourth order, we can obtain the diagrammatic expression of the flow equation for the self-energy $\Sigma_k=G^{-1}-\Gamma_k^{(2)}[\varphi=0]$ and for the four-point vertex function $\Gamma_k^{(4)}[\varphi=0]$ as in Fig. \ref{fig:flow_se} and in Fig.\ref{fig:flow_4pt}, respectively. 


\subsection{Flow of the four-point vertex in the vacuum}\label{subsec:flow_vac}
Before applying the f-FRG formalism with a vertex IR regulator to BEC of composite bosons, we first consider its flow in the vacuum. This situation can be realized by taking $\mu=-1/(2m a_s^2)-0^+$ and the limit $T\to 0$. In this limit, all the diagram which contains a closed loop of fermions vanishes automatically, since there are no antiparticles in non-relativistic physics. 
Therefore, the fermionic self-energy correction $\Sigma_k(p)$ vanishes since all the diagram in Fig.\ref{fig:flow_se} must contain a closed fermion loop: $\Sigma_k\equiv 0$.  

\begin{figure}[t]
\bea
\p_k\parbox{2.5em}{\includegraphics[width=2.5em]{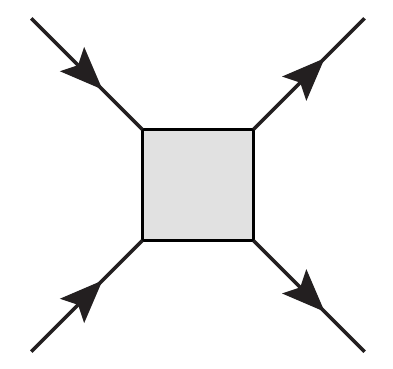}}&=&
\parbox{3em}{\includegraphics[width=3em]{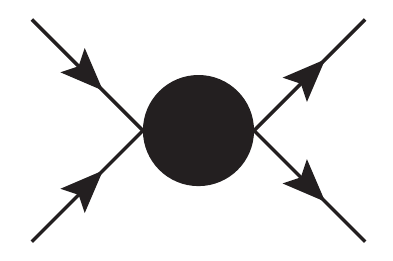}}+\parbox{5.5em}{\includegraphics[width=5.5em]{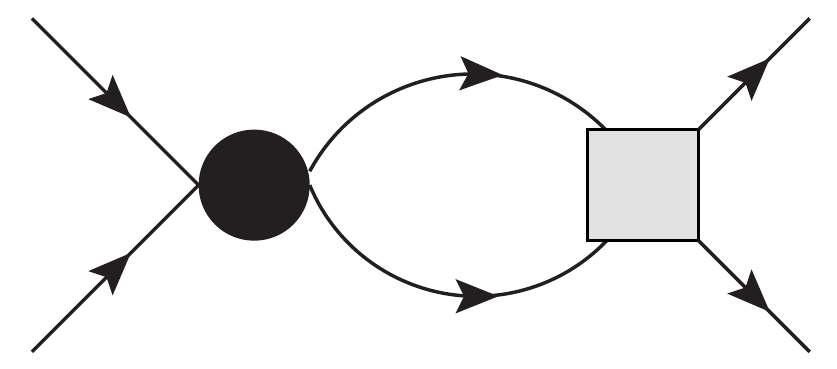}}+\parbox{5.5em}{\includegraphics[width=5.5em]{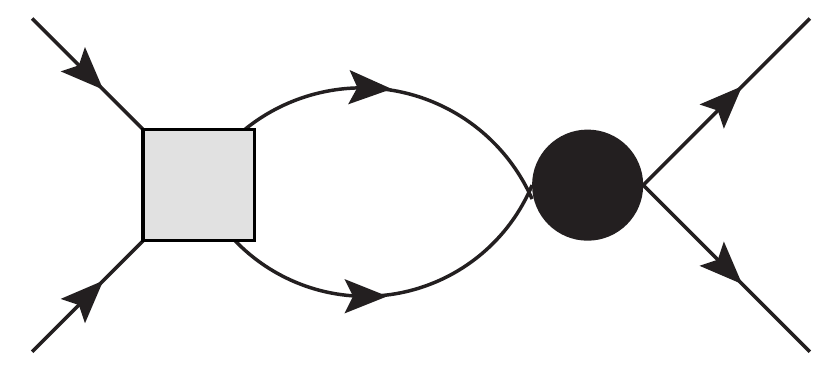}}+\parbox{8em}{\includegraphics[width=8em]{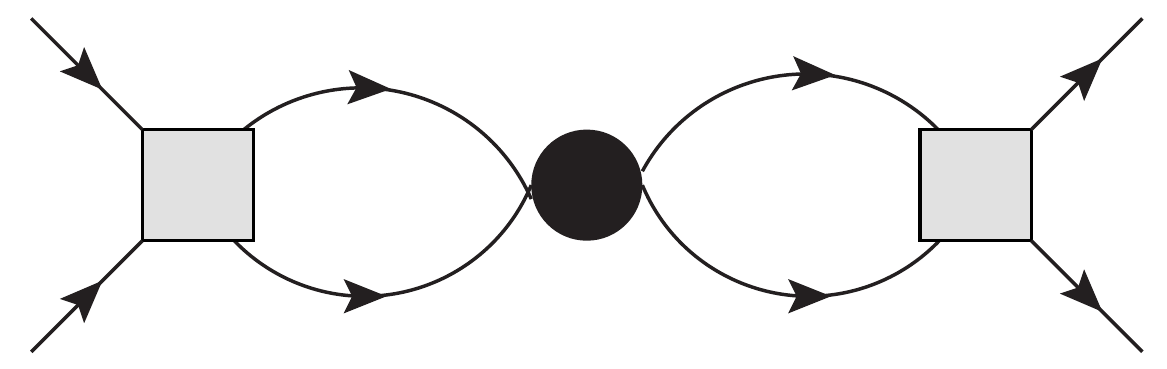}}
\nonumber
\eea
\caption{Flow equation of the four-point vertex function in the vacuum.}
\label{fig:flow_4pt_vac}
\end{figure}

Let us consider the f-FRG flow of four-point vertex function. For the flow equation of the four-point vertex function, the first three types of diagrams with particle-particle loops in the r.h.s. of Fig.\ref{fig:flow_4pt} survives in the vacuum limit and the other diagrams drop out. Therefore, its diagrammatic expression is greatly simplified and given in Fig.\ref{fig:flow_4pt_vac}. 
We denote the four-point vertex as $\Gamma_k^{(4)}(p;q,q')$, with $p$ the center-of-mass momentum and $q$ and $q'$ the relative momentum of in- and out-going particles, respectively. Then, the analytic form of the flow equation is given by 
\bea
\p_k \Gamma_k^{(4)}(p;q,q')&=&\p_k g_k(p)-\int_{l\le\Lambda}^{(T)}\left({(\p_k g_k)(p)\Gamma_k^{(4)}(p;l,q')\over G^{-1}({p\over 2}+l)G^{-1}({p\over 2}-l)}+{\Gamma_k^{(4)}(p;q,l)(\p_k g_k)(p)\over G^{-1}({p\over 2}+l)G^{-1}({p\over 2}-l)}\right)\nonumber\\
&+&\int_{l,l'\le\Lambda}^{(T)}{\Gamma_k^{(4)}(p;q,l)(\p_k g_k)(p)\Gamma_k^{(4)}(p;l',q')\over G^{-1}({p\over 2}+l)G^{-1}({p\over 2}-l)G^{-1}({p\over 2}+l')G^{-1}({p\over 2}-l')}. 
\label{vac01}
\eea
Since there are no explicit relative momentum dependence in the flow equation (\ref{vac01}) and in the initial condition at $k=\infty$, the solution does not have $q$ and $q'$ dependences either. From now on, we therefore denote $\Gamma_k^{(4)}(p)\equiv \Gamma_k^{(4)}(p;q,q')$.  We define a $p$-dependent function $M(p)$ by 
\be
M(p)=\int_{l\le \Lambda}^{(T)}{1\over G^{-1}({p\over 2}+l)G^{-1}({p\over 2}-l)}=\int_0^{\Lambda}{\diff^3\bm{l}\over (2\pi)^3}{1\over \bm{l}^2/m+\left(i p^0+\bm{p}^2/4m-2\mu\right)}, 
\label{vac02}
\ee
then the flow equation (\ref{vac01}) takes the form 
\be
\p_k \Gamma_k^{(4)}(p)=\p_k g_k(p) \left[M(p)\Gamma_k^{(4)}(p)-1\right]^2. 
\label{vac03}
\ee
We can integrate the differential equation (\ref{vac03}) with the initial conditions $\Gamma_{k=\infty}^{(4)}(p)=0$ and $g_{k=\infty}(p)=-g$ so as to find that 
\be
{1\over \Gamma_k^{(4)}(p)}={1\over g}-R_k(\bm{p})+M(p)=-\left(\Gamma_b^{(2)}(p)+R_k(\bm{p})\right). 
\label{vac04}
\ee
Therefore, the fermionic four-point vertex function $\Gamma_k^{(4)}(p)$ represents the inverse propagator of composite bosons when $k=0$, and the IR regulating function $R_k(\bm{p})$ correctly suppresses the low-energy bosonic excitations as we expected.

\subsection{Flow of the self-energy of f-FRG in the BEC limit}\label{subsec:flow_se}
Let us consider the many-body effect of the fermionic self-energy to find the number density of particles in the BEC limit. In this section, we do not take into account the many-body effect for the four-point function and approximate its flow equation by the one discussed in Sec.\ref{subsec:flow_vac}. This approximation corresponds to the particle-particle random phase approximation (PP-RPA), if the self-energy correction in the fermionic propagators in each diagram of Fig.\ref{fig:flow_4pt_vac} is neglected. 

\begin{figure}[t]
\bea
\p_k\parbox{4em}{\includegraphics[width=4em]{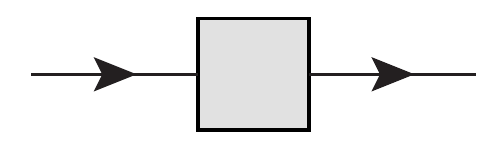}}&=&
\parbox{4em}{\includegraphics[width=4em]{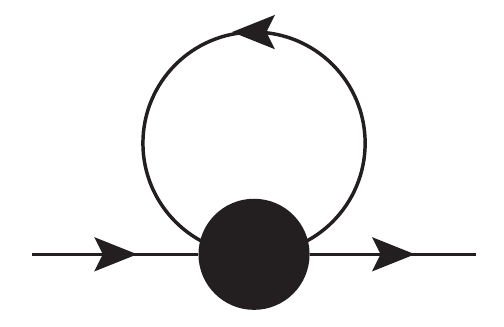}}+\parbox{6em}{\includegraphics[width=6em]{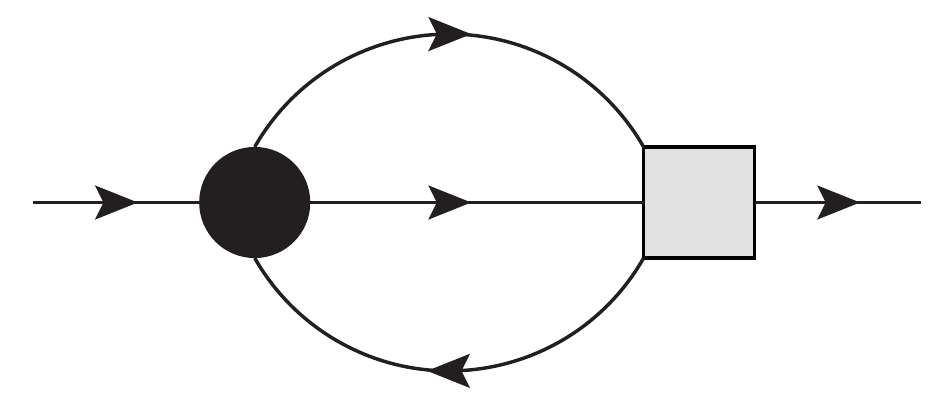}}+\parbox{6em}{\includegraphics[width=6em]{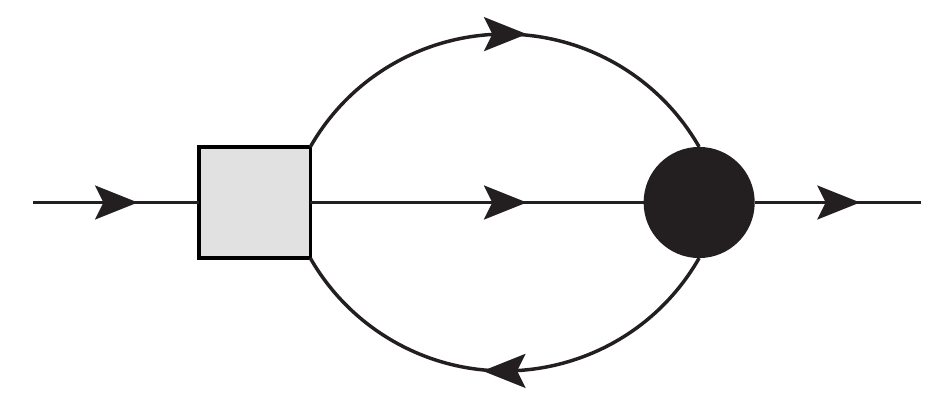}}+\parbox{7em}{\includegraphics[width=7em]{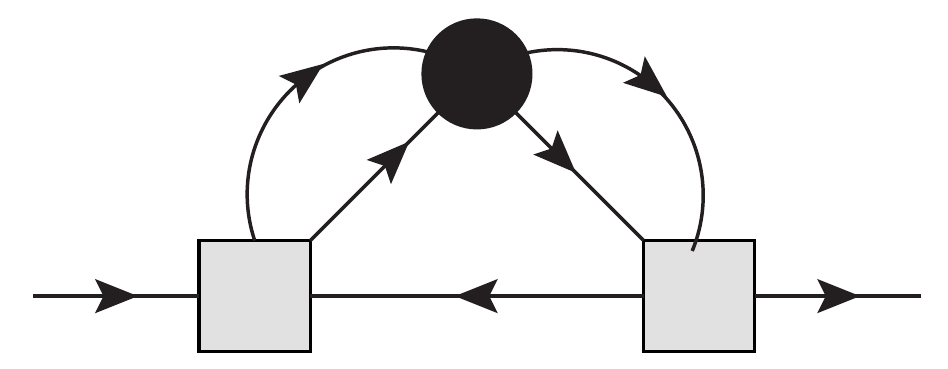}}
\nonumber
\eea
\caption{Approximate flow equation of the self-energy for a low-density system.  }
\label{fig:flow_se_lowdensity}
\end{figure}

Since we are interested in the low-density system, the flow equation for the self-energy in Fig.\ref{fig:flow_se} can be simplified as follows. Since each closed particle loop produces the particle number density, the diagrams only with a single closed loop must give the dominant contribution. Therefore, the fourth and fifth diagrams in the r.h.s. of Fig.\ref{fig:flow_se} are negligible for low-density systems, and the flow equation of the self-energy can be simplified as in Fig.\ref{fig:flow_se_lowdensity}. 
By substituting the solution $\Gamma_k^{(4)}(p)$ of the flow equation for four-point function in Fig.\ref{fig:flow_4pt_vac} into the flow of self-energy in Fig.\ref{fig:flow_se_lowdensity}, we can find that 
\be
\p_k \Sigma_k(p)=\int_{l}^{(T)}{\p_k \Gamma_k^{(4)}(p+l)\over G^{-1}(l)-\Sigma_k(l)}. 
\label{flow_se01}
\ee
By neglecting the self-energy in the r.h.s. of (\ref{flow_se01}), we can readily solve this differential equation. 
In this approximation, the solution of the flow equation (\ref{flow_se01}) is given by 
\bea
\Sigma_k(p)= \int_l^{(T)}{\Gamma_k^{(4)}(p+l)\over G^{-1}(l)} 
\simeq \int{\diff^3\bm{q}\over (2\pi)^3}{ {(8\pi/ m^2 a_s)} n_B\left({\bm{q}^2/ 4m}+\widetilde{R}_k(\bm{q})\right)\over i p^0+{\bm{q}^2/4m}+\widetilde{R}_k(\bm{q})-(\bm{q}+\bm{p})^2/2m-1/2ma_s^2}, 
\label{sol_se}
\eea
with $n_B$ the Bose-Einstein distribution function, $\mu=-1/2ma_s^2$, and $\widetilde{R}_k={8\pi\over m^2 a_S}R_k$. Details of the calculations can be found in Appendix \ref{supp:se}. 
This approximation can be justified by comparison of magnitude between $\Sigma_k$ and $\mu$. 
Even for small $k$ compared with the scale of temperature, the magnitude of the self-energy can be estimated as 
\be
|\Sigma_k(p)|\lesssim {1\over 2m a_s^2}\times (\sqrt{2m T}a_s)^3\times n_B(k^2/4m), 
\ee
which is much smaller than that of the chemical potential $|\mu|=1/(2m a_s^2)$ when $a_s\to 0^+$ as long as $n_B(k^2/4m)\sim 1$. 
Therefore, $\Sigma_k$ in the r.h.s. of (\ref{flow_se01}) is negligible for the most part of the flow in the deep BEC limit. Indeed, the critical value $k_c$ of this approximation would be estimated as $(k_c^2/2m)/ T\sim (\sqrt{2m T}a_s)^3\sim n a_s^3\ll 1$. 

In the following, instead of more sophisticated quantitative analysis, we discuss qualitative behaviors of the flow of f-FRG with the vertex IR regulator in the deep BEC limit. According to the above argument, we approximate the flow of the self-energy and the four-point vertex as 
\bea
\Sigma_k(p)&=&\int{\diff^3\bm{q}\over (2\pi)^3}{{(8\pi/ m^2 a_s)} n_B\left({\bm{q}^2/ 4m}+\widetilde{R}_k(\bm{q})-\mu_d\right)\over i p^0+{\bm{q}^2/4m}+\widetilde{R}_k(\bm{q})-(\bm{q}+\bm{p})^2/2m-1/2ma_s^2-\mu_d/2}, \\
\Gamma_k^{(4)}(p)&=&-{8\pi/m^2 a_s \over i p^0+\bm{p^2}/4m-\mu_d+\widetilde{R}_k(\bm{p})}. 
\eea
Here we introduced $\mu_d$ as $\mu=-1/2m a_s^2-\mu_d/2$. The Thouless criterion of the superfluid phase transition \cite{thouless1960perturbation} is given by the divergence of $\Gamma_{k=0}^{(4)}{(p)}$ at the zero center of mass momentum $p=0$, which implies that $\mu_d=0$, or $\mu=-1/2ma_s^2$. 
In order to determine the ratio between the critical temperature $T_c$ and the Fermi energy $\e_F$, we must calculate the number density $n$ of fermions. In this formulation, it is determined as 
\be
n=\int^{(T)}_p {-2e^{-ip^0 0^+}\over G^{-1}(p)-\Sigma_0(p)}. \label{eq:number}
\ee
When we expand the integrand in terms of $\Sigma_0(p)$, the main contribution of this integration comes from the term $G(p)\Sigma_0(p)G(p)$, which gives 
\be
n\simeq 2\int{\diff^3\bm{q}\over (2\pi)^3}n_B(\bm{q}^2/4m)={(2m T_c)^{3/2}\over \pi^2}\sqrt{\pi\over2}\zeta(3/2). 
\label{n_eq}
\ee
Details of this calculation is also given in Appendix \ref{supp:se}. 
We get the critical temperature $T_c/\e_F=0.218...$, which is nothing but the BEC transition temperature of free Bose gas. 

\subsection{Systematic improvement from the deep BEC limit}\label{subsec:improv}
In Sec.\ref{subsec:flow_se}, we derived an approximate solution of the f-FRG flow equation to describe BEC of free Bose gas in the deep BEC limit. 
In this subsection, we discuss the possibility of quantitative and systematic improvement of approximations to describe BEC of interacting Bose gas in the BEC side of the BCS-BEC crossover. 

Since each closed loop in a Feynman diagram produces a factor related to the number density, we would be able to construct a systematic expansion with a parameter $n a_s^3$ by analyzing the loop structure of diagrams. Its lowest order approximation is given in previous subsections and describes the BEC of free Bose gas. 
This kind of study will open a way to understand how thermodynamic properties can be different between the systems of elementary bosons and of composite bosons. 

As we take into account the higher order correction in terms of the number density, contribution from higher-point vertex functions must appear. Therefore, we must solve the scattering problem in the vacuum for the six- and eight-point vertex functions using the f-FRG flow equation as we have done for the four-point vertex in Sec.\ref{subsec:flow_vac}. 
As in the case of the FRG with two-point IR regulators, the flow equation may not be closed in its original form even for non-relativistic scattering problems. The flow equation of few-body physics, however, can be rewritten in a closed one by classifying the structure of higher-point 1PI vertices \cite{Floerchinger:2013tp,tanizaki2013flow}. 
The same proof can be done for the FRG with the vertex IR regulator, which will be reported elsewhere. 
Therefore, the atom-dimer and dimer-dimer scatterings are calculable within our framework of FRG, and their effects can be taken into account for many-body theories. 

In order to follow this program of including the effect of higher-point vertices, the numerical calculations of the flow equation would  be required. However, as we can see from the coefficient of the vertex IR regulator, $g_k=g^2R_k/(1-gR_k)$, each term of the flow equation can depend on the bare coupling $g$ and UV cutoff $\Lambda$. 
It is of great importance to develop a technique to perform the UV renormalization of our new flow equation not only for completeness of theoretical frameworks but also for numerical computation.

\section{Summary and Conclusion}\label{sec:summary}
We proposed a new formalism of fermionic FRG (f-FRG) which can describe the deep BEC limit of an interacting fermion system. In order to suppress low-energy one-particle excitations of dimers, we introduced a vertex IR regulator in the four-fermion interaction instead of a usual IR regulator in the fermion propagator. 
The exact evolution equation of f-FRG with the vertex IR regulator is derived and its structure is analyzed in details. 

This new formalism of f-FRG is applied to the two-component resonantly-interacting fermionic system with a positive scattering length. 
We first solve the flow of the four-point vertex function in the vacuum, and explicitly show that the four-point function is an IR-regularized propagator of a dimer. 
Since the low-energy excitations are suppressed appropriately, the fermionic self-energy correction turns out to be small compared with the chemical potential for the most part of the f-FRG flow in the deep BEC regime. 
This enables us to treat complicated momentum dependence of the self-energy by solving the flow equation with some plausible approximations, which is important to calculate the number density within the framework of fermionic field theories. The BEC critical temperature $T_c/\e_F=0.218$ of the free Bose gas is obtained based on f-FRG without bosonization. 

We also discussed the possible direction of improvement to go beyond the free Bose gas limit. Analysis of the loop structure of each diagram will provide a systematic approximation of the flow equation in the low-density limit.  
For such quantitatively sophisticated analysis, some numerical calculations of the flow equation is necessary, and it becomes important to rewrite the flow equation with the vertex IR regulator into a suitable form of numerical computations. 

As a concluding remark, let us discuss the possibility of f-FRG to describe the whole region of the BCS-BEC crossover. 
If we tried to describe the BCS side using f-FRG with the vertex IR regulator, the RG flow would be difficult to control due to the existence of low-energy fermionic excitation. 
In the BCS region, fermionic FRG with the IR regulator inside the fermion propagator gives a suitable description of the low-energy physics by taking into account the effect of Fermi surface. It enables to describe the BCS theory and its Gorkov and Melik-Barkhudarov correction, and to systematically exceed those conventional theories of superfluidity \cite{Tanizaki:2013doa,Tanizaki:2013fba}. 
In the BEC region, the vertex IR regulator controls the FRG flow so as to describe the BEC of composite particles, as discussed in this paper. 
Therefore, in order to describe the whole region of the crossover, combining these two techniques of fermionic FRG is promising since this make all kinds of low-energy degrees of freedom under control of the FRG flow. 
This work is now in preparation and will be reported elsewhere. 

\appendix
\section{Supplement of calculations in Sec.\ref{subsec:flow_se}}\label{supp:se}\newcommand{\tint}[1]{\int^{(T)}_{#1}\hspace{-.6em}}
We will show details of calculations for $\Sigma$ and $n$. For simplicity, we put $\mu=-1/2ma_s^2$ and use the unit $2m=1$. Analytic form of the approximate solution of the flow equation (\ref{flow_se01}) is given by 
\begin{eqnarray}
\Sigma_0(p)=\int_l^{(T)}{\Gamma_k^{(4)}(p+l)\over G^{-1}(l)}=
-{16\pi\over a_s}\tint{l}e^{-i l^0 0^+}
{1+\sqrt{1+{a_s^2\over 2}(il^0+ip^0+(\bm{p}+\bm{l})^2/2)}\over 
[i l^0 +\bm{l}^2+1/a_s^2][il^0+ip^0+(\bm{p}+\bm{l})^2/2]}. 
\label{nsr15}
\end{eqnarray}
Let us perform the Matsubara sum, which gives 
\begin{eqnarray}
&&{1\over \beta}\sum_{l^0}
{1+\sqrt{1+{a_s^2\over 2}(il^0+ip^0+{(\bm{p}+\bm{l})^2\over 2})}\over 
[i l^0 +\bm{l}^2+{1\over a_s^2}][il^0+ip^0+{(\bm{p}+\bm{l})^2\over 2}]}\nonumber\\
&=&
-{2n_B\left((\bm{p}+\bm{l})^2/2\right)\over ip^0+(\bm{p}+\bm{l})^2/2
-\bm{l}^2-1/a_s^2}
\label{nsr17}\\
&-&n_F(\bm{l}^2+1/a_s^2)
{1+\sqrt{{a_s^2\over 2}(ip^0+{(\bm{p}+\bm{l})^2\over 2}
-\bm{l}^2+{1\over a_s^2})}\over ip^0+(\bm{p}+\bm{l})^2/2-\bm{l}^2-1/a_s^2}\nonumber\\
&+&\int_{C_2}{\diff \omega\over 2\pi i}
{-1\over e^{\beta\omega}+1}
{1+\sqrt{1+{a_s^2\over 2}(-\omega+ip^0+{(\bm{p}+\bm{l})^2\over 2})}\over 
[-\omega+\bm{l}^2+{1\over a_s^2}][-\omega+ip^0+{(\bm{p}+\bm{l})^2\over 2}]},\nonumber
\end{eqnarray}
where $C_2$ is the contour wrapping around the branch cut $\{x+i p^0\in\mathbb{C}|x\ge (\bm{p}+\bm{l})^2/2+2/a_s^2\}$ of the squared root in the integrand. 
The second and third terms in the right hand of (\ref{nsr17}) are exponentially suppressed 
in the limit $\beta/a_s^2\to \infty$. 
Let us neglect these higher order corrections to get the formula corresponding to (\ref{sol_se}): 
\begin{equation}
\Sigma_0(p)={16\pi \over a_s}\int{\diff^3\bm{l}\over (2\pi)^3}
{2n_B \left( {(\bm{p}+\bm{l})^2/2}\right)
\over ip^0+(\bm{p}+\bm{l})^2/2-\bm{l}^2-1/a_s^2}. \label{nsr18}
\end{equation}

We can now evaluate the number density $n$ via the formula (\ref{eq:number}) using the self-energy given by (\ref{nsr18}): 
\begin{eqnarray}
n
&\simeq& \tint{p}(-2)G(p)\Sigma_0(p)G(p)\nonumber\\
&=&
-2\int{\diff^3\bm{p}\over (2\pi)^3}{\diff^3\bm{l}\over (2\pi)^3}{1\over \beta}\sum_{p^0}
{(32\pi /a_s)n_B \left( {(\bm{p}+\bm{l})^2/ 2}\right)
\over [ip^0 +\bm{p}^2+{1\over a_s^2}]^2
[ip^0+{(\bm{p}+\bm{l})^2\over 2}-\bm{l}^2-{1\over a_s^2}]}.\label{nsr19}
\end{eqnarray}
We here expand the full propagator in terms of the self-energy $\Sigma(p)$ in this expression and extract the dominant contribution. 
 The Matsubara sum gives 
\begin{equation}
{1\over \beta}\sum_{p^0}{1
\over [ip^0 +\bm{p}^2+{1\over a_s^2}]^2
[ip^0+{(\bm{p}+\bm{l})^2\over 2}-\bm{l}^2-{1\over a_s^2}]}
\simeq -{n_F\left( {(\bm{p}+\bm{l})^2/ 2}-\bm{l}^2-{1/ a_s^2}\right)
\over [{(\bm{p}+\bm{l})^2\over 2}-\bm{l}^2-\bm{p}^2-{2\over a_s^2}]^2}, 
\label{nsr20}
\end{equation}
where we again neglect exponentially small contributions in terms of $\beta/a_s^2$. 
We obtain that 
\be
n=
2\int{\diff^3\bm{P}\over (2\pi)^3}n_B\left({\bm{P}^2\over 2}\right)
\int {\diff^3\bm{q}\over (2\pi)^3}{ {32\pi\over a_s}n_F({\bm{P}^2\over 2}-({\bm{P}\over 2}-\bm{q})^2-{1\over a_s^2})\over \left[{(2\bm{q})^2/ 2}+{2/ a_s^2}\right]^2}. 
\label{nsr21}
\ee
Here we have changed integration variables so that $\bm{P}=\bm{p}+\bm{l}$ and 
$\bm{q}=(\bm{p}-\bm{l})/2$. 
Let us evaluate the integration over the relative momentum $\bm{q}$. 
Since we may approximate the Fermi distribution function as 
$n_F(-\bm{q}^2-1/a_s^2)\simeq 1$, the relative momentum integration gives one: 
$\int {\diff^3\bm{q}\over (2\pi)^3}
{ {32\pi/ a_s}\over \left[2\bm{q}^2+{2/ a_s^2}\right]^2}=1$. 
Therefore, the number density $n$ is given by (\ref{n_eq}). 

\section*{Acknowledgments}
The author wishes to thank Tetsuo Hatsuda for insightful suggestions and for carefully reading the manuscript. The author also thanks Gergely Fej\H{o}s for useful discussion. 
Y.T. is supported by JSPS Research Fellowship for Young Scientists. This work was partially supported by RIKEN iTHES project and by the Program for Leading Graduate Schools, MEXT, Japan.

\end{document}